# ADVANCED VEHICLE SAFETY and CONTENT DISTRIBUTION SYSTEM

Anita B[#1], Beena Sheril[*2], Ramesh B.E[#3]

[#] (M.Tech) CSE, SJM Institute of Technology, Chitradurga, India

[#] Lecturer, Department of Computer Science and Engineering
SJM Institute of Technology and Management, Chitradurga,India

[#] Asst.Prof, Department of Computer Science and Engineering
SJM Institute of Technology and Management, Chitradurga- India

[1]anitab0304@gmail.com
[2]beenasheril@gmail.com
[3]be.ramesh@gmail.com

*Abstract-* Advanced vehicle content distribution system (ACDS)is complemented by improved network connectivity with Mobile Network 3G, 4G network. Advanced content distribution system uses Access Points deployed along roadside. APs co-ordinate and collaborate to distribute content to vehicles in mobility. The infrastructure of deployed APs solves real time issues like predicting errors in movement, limited information shared to vehicles on movement due to limited resources. The advances vehicle content distribution system structures APs in to form a map which is considering the vehicle contact pattern which is analyzed by APs. The system is more effective by optimizing the network consumption by sharing prefectched data between APs. The process depends on APs storage, bandwidth and load on the origin APs which is connected to internet.  With the features Advanced System to distribute the content the system takes care of improving road safety, delivery accuracy and important content distribution.

## I. INTRODUCTION

To distribute Content to vehicular users through Access Points(APs) while wireless network access is emerging as a necessity to facilitate better road safety and enhance driving experience. The types of contents can include electronic newspapers, advertisements, road-situation reports, maps with traffic statistics, music or movie clips, etc. The cellular network (e.g., 3G or 4G) mainly adopted due to its ubiquitous availability, is experiencing an explosive growth in subscribers and in demands for multimedia contents, thus risking being pushed to its capacity limit.In this paper we are introducing ACDS to distribute content to vehicular user.

A typical architecture of WiFi-based vehicular content distribution system as illustrated in Fig 1. is made up of a network of interconnected APs, which are geographically deployed near the roads, running the customized protocols for cooperation and also equipped with local storage. The APs can communicate with each other through backhaul links to the Internet or via high-speed LAN.

Meanwhile WiFi-based Access Points (APs), as promising complement and augmentation to the cellular networks, have shown their feasibility in content distribution for vehicles . These APs can be RoadSide Units (RSUs) deployed intentionally by network service providers and government departments or Hot-spots that are installed in roadside shops or buildings and configured for public access. These APs are characterized by short-range coverage (hundreds of meters) relatively cheap and easy deployment and high data access rate (a theoretically 600 Mbps data rate in the latest IEEE 802.11n). A map which is deployed on the top of Aps which communicates each other. Data-origin servers are the content providers, providing vehicular users with both the. shared (popular) and also private contents.

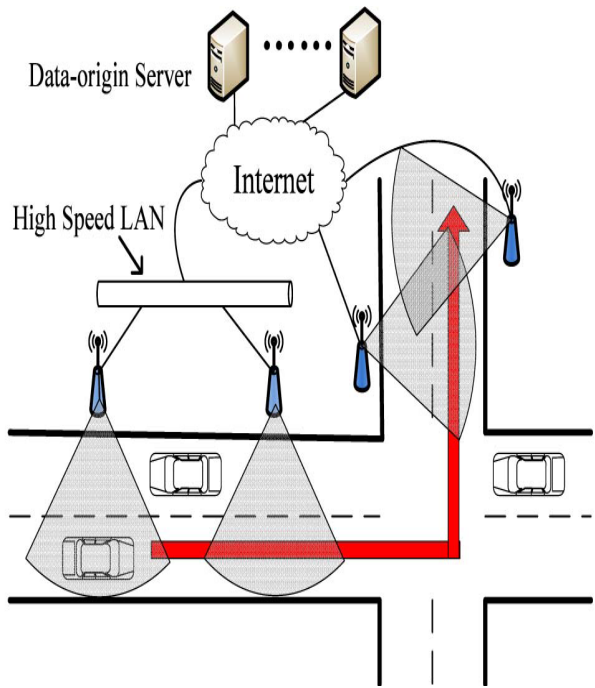

Fig 1. Architecture of vehicular content distribution system

## II. LITERATURE SURVEY

Previous work and proposed work will be considered

### A. EXISTING SYSTEM

The APs can communicate with each other through backhaul links to the Internet or via high-speed LAN. Data-origin servers are the content providers, providing vehicular users with both the shared (popular) and also private contents.

However, such network access scheme poses many challenges on the system design for effective content distribution to vehicles:

*DISADVANTAGES*: 1) The response latency of remote data-origin server on the Internet can waste the valuable contact duration, especially for the heavily loaded server and congested or long-delay route; 2) The wireless bandwidth between the AP and vehicles can be bottlenecked by the AP's backhaul path to the server on the Internet, considering that most of the WiFi-based access protocols are in the order of 10 Mbps while a median bandwidth of 5 Mbps of wired bandwidth to residents is reported.3) A single vehicle-AP contact duration is quite limited (typically tens of seconds) due to the fast vehicle speed and the AP's short coverage range, thus constraining the data transfer opportunities;

### B. PROPOSED SYSTEM

To solve the above challenges in vehicular content distribution, several recent works adopted prefetching technique that has a long history and is widely applied to computer architecture, CPU design and web accessing, etc. In these works, the requested data is prefetched onto the APs ahead along the driving trajectory of the requesting vehicle, which can then download the prefetched data with a high throughput when connection is established, without resorting to the remote server or being bottlenecked by the AP's backhaul link. Such apparently simple technique, however, requires careful design when it is applied onto large-scale and real systems. It will be very effective to deliver the data. *ADVANTAGES*: 1)Predicating errors in movement.2).prefetched data will be shared

## III. SYSTEM DESIGN

Fig. 2 shows all the probable to-be-contacted APs which are 3 hops ahead

(these APs are abbreviated as lookahead-APs) along the vehicle's path, in a tree-structure with transition probabilities being marked on the edges.

In general, A needs to consider prefetching on the APs within k-hops (k 1) ahead, usually more than just one hop, in order to give these APs ahead enough time to complete the prefetch before the requesting vehicle's arrival. For the selection of prefetching APs, we consider two cases in Fig. 2: 1) A gives notification of prefetching to all the lookahead-APs to get the maximum cache hit probability2) A issues notification of prefetching to B;F;... and J which is the predicted most probable vehicle-AP contact sequence; These two cases correspond to the two extremes of the spectrum and face disadvantages. The first case is suitable if the storage and backhaul connected ahead by the requesting vehicle, of prefetching (parts of) the uncompleted portion. We use Fig. 2 to show a high level view of how the advanced works from AP A's perspective. Here, AP A receives a request for file, say M, from an associated vehicle and then transmits data of M from either A's local storage or the remote server hosting M. At the same time, A Aps can be forecasts whether the download of M can be completed before the vehicle drives out of the coverage range. If not, A is responsible to select and notify those APs which may probably be bandwidth are sufficiently abundant, otherwise such excessive redundant prefetching may risk overwhelming these resources: the ejected prefetched data for the other sessions may degrade the overall performance. Moreover, the overwhelming traffic introduced into APs can bring down the throughput of the users currently being served due to the shared backhaul link.

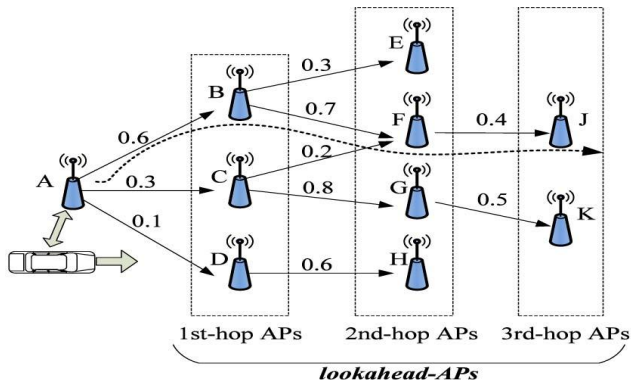

Fig. 2. High-level overview of operations from AP A's perspective. Transition probability is marked beside each edge. (dashed curve is the predicted most probably vehicle-AP contact sequence B ! F ! J).

The second case costs the least resources but has the worst robustness against mobility prediction error, which makes the prefetched data useless and thus wastes the invested resources (storage, backhaul bandwidth, etc.).

## IV. CONCLUSIONS

In this paper, we propose ACDS for vehicles through infrastructure APs. This system is designed to achieve efficient cooperation among the network of APs so that vehicular users can effectively utilize the opportunistically encountered and short-lived AP connections. A structure called map which is communicate with Aps is maintained distributedly on top of the APs, learning and predicting the potential vehicle-AP contacts. With the representative-based scheme, ACDS carefully selects the ones from the predicted set of lookahead-APs to perform prefetching in order to avoid overloading AP's backhaul link and ejecting the prefetched/cached data whose cost may overwhelm the prefetching benefits. ACDS distributes the contents encoded by network coding and to maximize the information utility under network coding, ACDS selects each piece of prefetched content based on the storage status (reflected by the metric rank-sum) of neighboring APs in the map. The simulation results under various scenarios prove the effectiveness of ACDS in many performance aspects.

ACKNOWLEDGMENT

Special thanks to SJM Institute of Technology for helping us in our work and supporting in research..Vishveshwaraya Technological research center also helping us in this work.

REFERENCES

[1]. Enabling Efficient WiFi based Vehicular Content Distrbution by DaZhang, Member, IEEE, and Chai Kiat Yeo

[2] . Y. Huang, Y. Gao, K. Nahrstedt, and W. He, "Optimizing File Retrieval in Delay-Tolerant Content Distribution Community," Proeb c. IEEE 29th Int'l Conf. Distributed Computing Systems (ICDCS), pp. 308-316, 2009.

[3]. B. Chen and M. Chan, "MobTorrent: A Framework for Mobile Internet Access from Vehicles," Proc. IEEE INFOCOM, pp. 1404- 1412, 2009.

[4]. D. Zhang and C.K. Yeo, "A Cooperative Content Distribution System for Vehicles," Proc. IEEE Global Telecomm. Conf. (Globecom '11), 2011.

[5]. M. Li, Z. Yang, and W. Lou, "Codeon: Cooperative Popular Content Distribution for Vehicular Networks Using Symbol Level Network Coding," IEEE J. Selected Areas in Comm., vol. 29, no. 1, pp. 223-235, Jan. 2011.

[6]. Prefetching : By Steve VanderWiel David J. Lilja